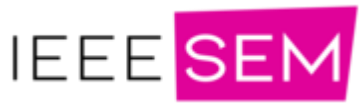

# Cyber Security Requirements for Platforms Enhancing AI Reproducibility*

**Polra Victor Falade, Nigerian Defence Academy**

¹Cyber Security Department, Nigerian Defence Academy, Kaduna, Nigeria
Email: pvfalade@nda.edu.ng

## ABSTRACT

Scientific research is increasingly reliant on computational methods, posing challenges for ensuring research reproducibility. This study focuses on the field of artificial intelligence (AI) and introduces a new framework for evaluating AI platforms for reproducibility from a cyber security standpoint to address the security challenges associated with AI research. Using this framework, five popular AI reproducibility platforms—Floydhub, BEAT, Codalab, Kaggle, and OpenML—were assessed. The analysis revealed that none of these platforms fully incorporates the necessary cyber security measures essential for robust reproducibility. Kaggle and Codalab, however, performed better in terms of implementing cyber security measures covering aspects like security, privacy, usability, and trust. Consequently, the study provides tailored recommendations for different user scenarios, including individual researchers, small laboratories, and large corporations. It emphasizes the importance of integrating specific cyber security features into AI platforms to address the challenges associated with AI reproducibility, ultimately advancing reproducibility in this field. Moreover, the proposed framework can be applied beyond AI platforms, serving as a versatile tool for evaluating a wide range of systems and applications from a cyber security perspective.

## 1 INTRODUCTION

IN contemporary scientific research, the significance of achieving reproducibility has become increasingly evident, particularly in computational research [1]. Reproducibility, the ability to replicate experimental outcomes and validate findings, stands as a fundamental cornerstone of scientific progress. This need for reproducibility is exceptionally pronounced in the realm of artificial intelligence (AI) and machine learning due to the intricate challenges posed by information security [2].

AI and machine learning have transcended the confines of academia and research labs to become integral components of modern society. They power essential applications across various domains, from healthcare to finance, making the verification of research findings critical [3]. Deploying AI systems without rigorous validation can have profound consequences, impacting the lives and well-being of individuals globally.

Victoria Stodden's comprehensive research compendia [4] elucidate the multifaceted nature of achieving reproducibility in research. The core lies in providing meticulous, all-encompassing information, including complete datasets, pseudocodes, algorithms, source codes, auxiliary materials, research papers, methodologies, and environmental factors. While this approach may pose minimal challenges in smaller-scale or clinical research, the landscape changes significantly when we enter the intricate realm of AI [5].

The unique challenge in AI research is its propensity to intersect with domains containing sensitive data, such as patient health records in healthcare settings. This data is not only highly confidential but also subject to stringent government regulations and legal frameworks [5]. Consequently, achieving reproducibility in AI research is not just an academic exercise; it's an imperative with profound implications for privacy, security, and accountability.

This research paper delves into the intricate relationship between cybersecurity and AI reproducibility, exploring the crucial requirements and safeguards needed to navigate this evolving landscape. As AI continues its integration into society's core functions, the assurance of reproducibility becomes intrinsically linked to our ability to responsibly and securely harness these transformative technologies.

## 2 REPRODUCIBILITY

Reproducibility has become a critical requirement in scientific research in recent years [6], especially when research is conducted across multiple institutions, and sites, or involves complex protocols, expensive machinery, and numerous staff to generate data [7]. Reproducibility ensures transparency and validates the research, but there is no universally accepted definition in the research community [7]. Goodman's definition of reproducibility, based on method reproducibility, result reproducibility, and inferential reproducibility, is one of the most commonly cited [8]. Another categorization of reproducibility, based on reproducibility under the same condition, reproducibility under variant conditions, and reproducibility in multi-test studies, has been proposed by Belz et al. [9].

Reproducibility is often confused with other similar concepts such as replicability, reliability, robustness, repeatability, generalization, and reuse [7], [8], [10]. Reproducibility, replicability, and reuse have been defined as distinct concepts by Baker et al. [7]. Reproducibility refers to the re-enactment of a study by a third party using the original set-up, data, and methodology of analysis [7]. Replicability, on the other hand, involves a more general re-enactment of results using the same analytical method on different datasets [7]. Re-use is the possibility of using the results beyond the original research context, both within and outside the scientific discipline [7]. However, Chris Drummond [11] takes a different stance, contending that replicability is a waste of time due to his unique definition of the term, which is contrary to other recent studies.

In the field of computer science, Victoria Stodden [12] has defined reproducibility as a means of organizing computational research to enable authors and readers of publications to verify reported results. Stodden identifies three categories of reproducibility based on the recorded information about the research: computational, empirical, and statistical reproducibility. Computational reproducibility involves keeping track of all information on the computational aspect of the research, such as the data, metadata, code, software, tools, and environment used, as well as the methods employed. Empirical reproducibility documents non-computational actions taken during the research process [12]. A research project is deemed to be reproducible if its results can be replicated by others given that the original code, data, and documentation are available [13].

To ensure that computational research is reproducible, certain minimum requirements must be met. These requirements, as outlined by Goodman [8], include the documentation of the dataset, metadata, code, and software used. Further to this, Victoria Stodden [12] has defined research compendia as a collection of artefacts that support the claims made in a research article, inclusive of the article itself. In a broader sense, this involves including several components, such as the research paper, data, experiment, results, and auxiliary material [4]. The inclusion of these components requires providing detailed descriptions of the data, code, methodology, and computing environment used, as well as high-resolution documentation of any results produced.

While reproducibility is essential for validating and verifying scientific research, replicability remains a critical goal of scientific research. However, replicability is often costly and not typically published in top-tier journals [13]. In contrast, reproducibility is a crucial aspect of scientific research that involves using the same data, methods, and documentation as the original research to obtain the same results, making it more attainable than replicability. The provision of such detailed information has several benefits in making research reproducible [8], [10], [13].

### 2.1 Reproducible AI Research

Russell et al. [3] contend that the primary emphasis of AI research should be on augmenting the competence of AI systems while also exploring methods to optimize the benefits that AI can offer to society. This involves ensuring that AI systems operate in a manner consistent with human objectives and do not behave in ways that are detrimental to humanity.

In their publication, Russell et al. [3] have categorized the priorities for AI research into two distinct categories: short-term and long-term priorities. The short-term priorities encompass several research areas in computer science, including the development of robust AI systems through verification, validity, security, and control. Additionally, there are concerns related to law and ethics such as privacy, policy questions, professional and machine ethics, liability and law for self-driving cars, and the deployment of autonomous weapons [14]. Finally, the optimization of the economic impact of AI, including labour market forecasting, market disruption, policy for managing severe effects, and financial measures, are also identified as short-term priorities [3].

On the other hand, the long-term priorities entail research on Artificial General Intelligence (AGI) and Artificial Super Intelligence (ASI). This research area also includes verification, validity, security, and control [3].

It is essential to investigate various aspects of AI research such as verification, validity, security, and control, which are equally important for both short-term and long-term research priorities. This paper aims to contribute systematically to both AI research priorities. Verification research involves ensuring that the AI system satisfies the minimum standard requirements and addresses the question of whether the system was constructed appropriately. Validity research, on the other hand, addresses the issue of whether the system behaves appropriately according to its intended purpose, thus avoiding unexpected or undesired behaviour. Security research focuses on protecting the AI system from malicious actions by both authorized and unauthorized users. Lastly, Control research allows humans to manipulate AI systems reasonably, especially in cases where misbehaviour arises [3]. Verifying and validating AI research emphasizes the importance of reproducible AI research in society.

The practice of publishing scientific findings arose from the recognition that withholding information impedes scientific progress [15]. However, published papers have inherent limitations in terms of the amount of information they can contain, making transparency and reproducibility challenging in computational research [15],[1]. The culture of knowledge-sharing and the increasing emphasis on reproducibility has led to the development of online tools that facilitate the sharing of research ideas and collaborations on a global scale [15],[1]. These platforms have the potential to enhance the reproducibility of research [1], but the traditional research paper remains a vital component of research documentation [4].

These challenges have led to the development of various platforms and tools that are specifically designed for AI and machine learning research. AI platforms designed for reproducibility were created to facilitate the exchange of large datasets, codes, workflows, methodologies, and findings within the context of AI and computational research.

Isdahl and Gundersen [1] introduced the concept of ML (Machine Learning) platforms for reproducibility (AI platforms), which they defined as software, cloud solutions, or environments that offer more advanced features than libraries and are used to facilitate the development of machine learning experiments. Through these platforms, researchers can more effectively document, track, and share their datasets, codes, workflows, and other essential details needed for reproducibility. Moreover, these platforms enable collaboration and contribution among researchers, as well as provide opportunities for learning and competition. Several examples of these platforms for machine learning reproducibility include OpenML, Kaggle, WaveLab, Google Cloud ML, Azure ML, BEAT, Floydhub, MLflow, studioML, Polyaxon, Kubeflow, Amazon Sagemaker, CometML, and Codalab [1]. Some of these platforms are also equipped with the capability to link to repositories.

In a broader sense, Victoria Stodden [12] proposed several factors that repositories should consider in ensuring reproducibility in AI and computational research. These factors encompass the provision of research compendia, versioning and the use of persistent unique identifiers for research artefacts, appropriate metadata and documentation, interoperability, open access to artefacts, open licensing for research artefacts, data and artefact ownership, confidentiality in data and analysis code, handling of extremely large data sets and code, and fostering industry collaborations. However, it is noteworthy that Stodden's suggestions did not explicitly address security concerns.

However, while these platforms have the potential to enhance the reproducibility of research, achieving full reproducibility in AI research remains a challenge. Due to the unique security challenges associated with AI and machine learning investigations. In AI research, ensuring reproducibility extends beyond simply establishing a platform for sharing, collaboration, and contribution. It necessitates creating provisions to address privacy concerns, intellectual property theft, and competitiveness - issues that are inherent to and almost exclusive to AI research [5]. This is because AI research may entail handling sensitive information subject to strict government regulations and laws. Hence the need to carefully consider the cyber security requirements of these AI platforms enhancing AI research reproducibility.

While reproducibility is crucial for AI research, as with other small and clinical research, it faces unique challenges that are specific to AI research in addition to general reproducibility challenges. One such challenge is the sensitivity of the raw data set, which is often sensitive data, such as healthcare data. Sharing this data with other researchers or the public can raise privacy concerns. Another challenge is that AI involves big data, and the data required to train the machine is often extensive, making the cost of collection and storage a challenge. Finally, researchers may be hesitant to share their work due to concerns about intellectual property theft [5]. These unique challenges related to privacy concerns and intellectual property theft in AI research are security issues that this paper aims to address. By solving these security issues associated with AI research reproducibility, we intend to enhance AI reproducibility.

## 3 METHODOLOGY

This research commences by proposing a set of cybersecurity attributes aimed at enhancing AI reproducibility. These attributes serve a dual purpose: ensuring that the essential prerequisites for reproducibility are met and addressing the security challenges commonly associated with AI research.

Following this, we introduce a comprehensive framework, which we subsequently employ to evaluate a selection of AI reproducibility platforms. The primary aim is to gauge how well these platforms incorporate the aforementioned cybersecurity attributes. This assessment plays a crucial role in determining the platforms' ability to promote AI reproducibility securely.

In the concluding phase of this study, we draw conclusions and recommendations based on the analysis outcomes. These recommendations are designed to assist diverse researchers in making informed decisions when selecting platforms that align with their specific preferences and needs. By tailoring platform recommendations to various use-cases, this research equips researchers with valuable guidance for navigating the ever-evolving landscape of AI reproducibility.

### 2.3 Cyber Security Properties

There are four main cybersecurity properties of a secure system: security, privacy, trust [16], and usability [17]. These properties have sub-factors that are interdependent and sometimes conflict with each other. For example, strong passwords ensure security but can make it difficult to remember, while providing too much information for verification purposes can violate privacy. It is necessary to have well-defined sub-factors to fully understand each security property.

However, there is a lack of uniformity in the definitions and naming of security properties in practice and academia. To address this issue, Michael Bitzer et al. [18] conducted an extensive literature review of security properties and identified several properties that can be used to ensure information security governance. They also provided more comprehensive definitions based on the existing definitions they found. This comprehensive list of information security properties helps achieve uniformity in in-formation security and also enables organizations to allocate resources more appropriately and ensure effective information security governance. The comprehensive list of information security properties includes accessibility, accountability, accuracy, admissibility, anonymity, auditability, authenticity, authorization, availability, confidentiality, consistency, human safety, integrity, intervenability, non-repudiation, possession, privacy, pseudonymity, reliability, responsibility, survivability, timeliness, transparency, trustworthiness, unobservability, and validity, as well as meta characteristics such as cost, effectiveness, efficiency, ethicality, suitability, usability, and utility [18]. This paper uses this comprehensive list and definitions of information security properties as sub-factors to security, privacy, usability, and trust as shown in Table 1, to provide a better understanding of each cyber security property of a secure system and application.

TABLE 1
CYBER SECURITY PROPERTIES WITH THEIR SUB-FACTORS AND CORRESPONDING DEFINITIONS

| Cyber Security Properties | Sub-factors | Definitions |
|---|---|---|
| Security | Authorization | "Act of determining whether an entity is allowed to perform an activity on a resource" [18] |
| | Auditability | "Ability to conduct persistent, non-bypassable monitoring of all actions performed by entities within the system" [18]. |
| | Authentication | "An act of validating a user's identity" [18] |
| | Access control | Access control determines who is allowed to use data and how [18]. |
| | Possession | "Resources are under the control and ownership of authorized entities" [18] |
| Privacy | Anonymity | "Ability of an entity to not be identified or at least undistinguished among another group of entities, if required" [18]. |
| | Pseudonymity | "Ability to use a resource without disclosing its entity identity, but can still be accountable for that use" [18]. |
| | Confidentiality | Protection of proprietary information from unauthorized disclosure or misuse" [18]. |
| | Unobservability | "The ability of an entity to use a resource without others being able to observe that the resource is used" [18]. |
| Usability | Accessibility | "Resources can be accessed and used by all authorized entities in a timely and reliable manner" [18]. |
| | Availability | "Resources can be accessed and used by all authorized entities in a timely and reliable manner" [18]. |
| | Timeliness | "Ensuring that necessary resources are available quickly enough when needed" [18]. |
| | System survivability | "Ability to maintain resource availability despite adverse circumstances" [18]. |
| | Intervenability | "Data subjects are effectively granted their rights to notification, information, rectification, blocking and erasure at any time and the controller is obliged to implement appropriate measures" [18] |
| | Human safety | "Safety of anyone dependent on the satisfactory behaviour and proper use of resources" [18]. |
| | Responsibility | "Handling the development of events in the future in a particular sphere" [18]. |
| Trust | Authenticity | "Authenticity is the property of an entity to be correct and genuine, to have the ability to be trusted, to have a verifiable identity, and to demand the same from other entities as well" [18]. |
| | Accountability | "Accountability is the property of a system to trace the actions of an entity and hold it uniquely responsible for its actions" [18]. |
| | Consistency | "Entities do what they are expected to do" [18]. |
| | Non-repudiation | "The ability of a system to prove the occurrence or non-occurrence of actions" [18]. |
| | Trustworthiness | "The ability of an entity to verify identity and establish trust in a third |

| | |
|---|---|
| | party" [18]. |
| Reliability | "Consistency in the intended behaviour and results" [18]. |
| Admissibility | "State in which the status of the data is acceptable or lawful" [18]. |
| Accuracy | "Data is free of errors and has the value that the affected entities expect" [18]. |
| Integrity | "Guarantee that all assets are functioning correctly and as intended" [18]. |
| Transparency | "The ability of a data subject, system operators, and supervisory authorities to understand how data is collected and processed for which purpose, as well as who is legally responsible" [18]. |
| Validity | "Information is up to date and has not been superseded by another" [18]. |

## 3.2 Proposed Framework for Analysis

The proposed framework serves as the foundation of this research, developed around four pivotal properties essential to secure systems and applications: security, privacy, usability, and trust, collectively referred to as SPUT (Security Privacy Usability Trust). The motivation behind this framework is to facilitate the comprehensive analysis of machine learning (ML) platforms in the context of reproducibility. It is worth noting that while this framework is tailored for AI platforms, its applicability extends beyond this domain to encompass various systems and applications.

In contrast to the traditional Confidentiality, Integrity, Availability (CIA) triad, which predominantly focuses on data security, the SPUT framework embraces a more holistic approach. The CIA triad's limitations stem from its exclusive emphasis on data security, thereby neglecting the broader spectrum of security considerations inherent in complex systems and applications.

In any system or application, safeguarding both user and data security is paramount. The users must have a certain level of assurance regarding their own safety and the security of the data they interact with. To achieve a nuanced understanding of the SPUT properties, we bifurcate each property into two categories: data security and user security, data privacy and user privacy, data usability and user system usability, and finally, data trust and user trust. This bifurcation recognizes that diverse individuals harbor distinct security expectations from various systems and applications, necessitating tailored approaches to cater to both data and user aspects.

Consider a scenario where a survey participant is engaged. In this context, the participant's foremost concern may revolve around safeguarding their user privacy rather than data privacy. The rationale behind this lies in the inherent nature of surveys, where responses are essential to fulfill the survey's purpose. Concealing responses would undermine the very essence of gathering information. However, preserving personal identity becomes paramount under these circumstances, constituting a facet of user privacy protection.

In contrast, envision a university application designed to disseminate critical information concerning lectures, exams, and timetables. Here, the emphasis shifts toward data trust as a pivotal security property among others. Students engaging with this application necessitate a high degree of confidence in the accuracy and reliability of the information it provides. Data trust, in this context, plays a crucial role in ensuring that students can rely on the application's data with certainty.

These examples underscore the dynamic nature of security considerations, wherein the specific context and objectives of a system or application dictate which facets of security, be it user privacy or data trust, assume precedence. Tailoring security measures to suit these distinct requirements is essential for ensuring the overall security and functionality of diverse systems and applications.

To establish sub-factors for the security of both data and users, we systematically classify various information security properties, sourced from a diverse array of literature. The decision to employ existing information security properties, specifically those comprehensively defined by Michael Bitzer et al. [18], stems from the desire to maintain consistency and clarity. This exhaustive literature review offers a unified and up-to-date perspective on information security properties, enhancing the framework's robustness. It is important to note that we introduce two additional security properties, namely authentication [19] and access control [20], to address specific requirements of our framework.

With a critical examination of each information security property's definitions, we categorize them as sub-factors under the umbrella of the four SPUT properties, ensuring that each property remains exclusive to its designated category. This classification extends further to distinguish between the security of data and user aspects within the SPUT framework.

The versatility of this framework shines through its applicability across a spectrum of systems and applications, serving various purposes. It is expressly designed to be accessible to a wide range of users, not exclusively catering to security experts. By eliminating the need for in-depth knowledge of internal code structures, this framework evaluates the services, features, and functionalities offered by different systems and applications, thereby ensuring its utility for a broad user base.

In practical application, the framework employs questions mapped to each property to simplify the understanding of complex security principles, enabling rapid assessment. Responses are categorized as 'Yes' (indicating provision of the security property), 'No' (indicating a lack of provision), or 'N/I' (No Information), serving as a valuable tool for users to ascertain the presence of security properties in systems and applications.

Moreover, before implementing the framework, it is essential to precisely define the terms "user" and "data," as these properties can be applied differently based on the specific context. For instance, considering confidentiality as a property, it is vital to specify whether it pertains to the confidentiality of personal information or user-uploaded data or other data utilized within the system.

Table 2 presents the outlined framework designed for the analysis process. In the forthcoming chapter, we leverage this proposed framework to assess selected AI platforms for reproducibility, putting it to the test and advancing the cause of AI reproducibility in the process

TABLE 2
PROPOSED FRAMEWORK BASED ON THE SPUT OF A SECURE SYSTEM AND APPLICATION

| Security Properties (SPUT) | Category | Sub-factors | Requirement Questions |
|---|---|---|---|
| **Security** | **User security** | Authorization | Are there policies that define what a user is allowed to do or not do on the system? |
| | | Auditability | Does the system keep a log of the activities by every user? |
| | | Authentication | Are users authenticated into the system? |
| | **Data security** | Access control | Are there permissions and privileges governing the use of data? |
| | | Possession | Are data assigned to owners? |
| **Privacy** | **User privacy** | Anonymity | Can a user hide identity while using the system? |
| | | Pseudonymity | Can a user hide identity while using data, but the log will record the activities? |
| | **Data privacy** | Confidentiality | Can data be stored and used privately? |
| | | Unobservability | Can data be used without other users' knowledge? |
| **Usability** | **Data usability** | Accessibility | Are data accessed from multiple devices whenever needed? |
| | | Availability | Are data available when the user needs it? |
| | | Timeliness | Are data accessed timely? |
| | **User system usability** | System survivability | Is there a provision for data to be preserved after system crashes? |
| | | Intervenability | Can users configure their accounts to fit them? |
| | | Human safety | Will users get in trouble for using data, maybe legal implications? |
| | | Responsibility | Is the system designed to handle future events? |
| **Trust** | **User Trust** | Authenticity | Can the system verify users' identities to be genuine? |
| | | Accountability | Does the system keep track of what users do and use on the system? |
| | | Consistency | Does the system detect unexpected behaviours from users? |
| | | Non-repudiation | Can a user deny actions carried out on the system? |
| | | Trustworthiness | Can users verify the identity of other users to establish trust with a new/strange user? |
| | | Reliability | Can the activities of the users over time be verified to determine if the user is genuine? |
| | **Data Trust** | Admissibility | Can the status of data be confirmed to be lawful? |
| | | Accuracy | Does the system analyze the data to be sure they are free from errors? Or does the system |

|  |  |
|---|---|
|  | detect errors in data? |
| Integrity | Can the integrity of data be verified to be free from unauthorized modification? |
| Transparency | Is there detailed documentation of information about how to use each data? |
| Validity | Can data be confirmed to be still valid or out of use? |

### 3.3 Selected Platforms for Analysis

In this paper, five AI platforms for reproducibility have been selected for analysis: Floydhub, BEAT, Kaggle, Codalab, and OpenML. Following an analysis of approximately thirteen ML platforms for reproducibility, as conducted by Isdahl and Gundersen (Isdahl & Gundersen, 2019), it was concluded that Floydhub and BEAT provide the strongest support for reproducibility, with Codalab and Kaggle being their close competitors. This paper aims to scrutinize these four AI platforms from a security perspective, utilizing the proposed framework. The objective is to assess whether these AI platforms implement the security properties that are crucial for supporting reproducibility or not. It is worth noting that OpenML was selected randomly for inclusion in this analysis.

It is important to clarify that the authors of this paper do not possess prior experience working with any of the five selected AI platforms. Consequently, the analysis of these platforms in this paper relies on the publicly available documentation for each AI platform. While combining documentation-based analysis with practical experience is ideal, it is recognized that documentation may not always provide a complete and accurate portrayal of all aspects of these platforms. Developers or the organizations that own these platforms may inadvertently or intentionally omit certain critical details in their documentation.

### 3.4 Methodological Shortcomings

The methodology employed in developing this framework and conducting the analysis presents several notable challenges and limitations. Firstly, the primary challenge was to propose a novel framework for systematically evaluating AI platforms from a security perspective. This was a daunting task due to the absence of prior systematic approaches in this area. The framework's design aimed for versatility, intending to be applicable to a wide array of systems and applications. However, the inherent diversity among systems and applications poses difficulties in linking each security property to a single, universally applicable parameter. Different systems may implement the same properties through distinct modalities, making it challenging to establish a one-size-fits-all approach. For instance, ensuring dataset confidentiality can be achieved through various modalities, such as restricting public access or offering options for dataset privacy. These modalities serve different roles in determining security levels, but for the paper's scope, the focus is solely on the presence or absence of such modalities, not their specific roles. While it's not impossible to conduct a quantitative or semi-quantitative analysis, the questions employed in this paper were chosen to accommodate the framework's general and adaptable nature (Author, Year).

Another limitation pertains to the method of data collection for this analysis. Relying solely on documentation was found to be insufficient in providing answers to all the questions raised by the framework. Ideally, combining documentation with the experiences of other users could have yielded more comprehensive results. However, due to the evolving nature of the framework throughout the paper's development, it was challenging to plan for such an approach. The inclusion of a third option, 'N/I' (No Information), alongside 'Yes' and 'No', was deemed necessary to avoid incorrect assertions that a platform lacks a certain property when information might be missing or undocumented.

The property of usability presents a unique challenge. Assessing usability accurately proved difficult, especially in the absence of personal experience using the different platforms. While datasets' availability was consistently addressed in all platforms, determining factors like accessibility and timeliness posed challenges, as they may vary considerably depending on user-specific contexts.

These limitations and challenges underscore the complexity of systematically evaluating AI platforms from a security perspective and emphasize the need for a well-rounded and comprehensive approach that combines diverse data sources and perspectives.

## 4  PROPOSED CYBER SECURITY PROPERTIES FOR ENHANCING AI REPRODUCIBILITY

The notion that all cyber security properties are universally applicable is a misconception. While cyber security properties are critical for protecting data and systems, they can also conflict with each other. Therefore, it is essential to determine which security properties are necessary for a particular system or application. For instance, anonymity can be useful in raising concerns in 'MySurreyVoice', a system designed for students at the University of Surrey. Anonymity allows students to express their opinions without fear of being judged or criticized by others. It also encourages them to speak up and express their concerns without revealing their identities. Similarly, anonymity is applicable in surveys, where participants usually prefer not to reveal their identities. However, anonymity may not be required in a Zoom video conference meeting where attendance must be taken, and participants cannot hide their identity. Therefore, it is crucial to identify the cyber security properties necessary for a specific system or application based on the functionality of the system.

Cyber security properties conflict with each other, therefore the need to find a trade-off of properties that will enhance the reproducibility of AI research through the AI platforms. The cyber security properties are proposed in two categories; those that will solve the security issues related to AI research and those that will ensure that the principles of reproducibility are upheld.

## 4.1 Cyber Security Properties that Mitigate Security Concerns Associated with AI Reproducibility

Section two of this paper identified security-related challenges to AI reproducibility, particularly those related to information security. These challenges include privacy concerns regarding the use of sensitive data in AI research, as well as the risk of intellectual theft and competitiveness within the field. Additionally, the possibility of malicious actors gaining access to AI datasets is a significant concern. Based on these challenges, the following security properties are recommended:

A. Privacy of sensitive datasets concerns

The use of sensitive datasets on a platform for machine learning model training requires careful consideration of their security, confidentiality, and admissibility. Implementing security measures such as confidentiality and access control, as well as ensuring admissibility, can help to address these concerns and prevent legal issues associated with the unauthorized use of sensitive data.

*Confidentiality(Privacy)*

The use of sensitive datasets, such as patients' healthcare data, for training machine learning models raises significant legal and ethical concerns. Unauthorized use or disclosure of personally identifiable information is a violation of privacy laws, such as the General Data Protection Regulation (GDPR). It is crucial to consider several factors when using such sensitive datasets on a platform to ensure their security, confidentiality, and admissibility. Confidentiality is a security property that provides privacy for sensitive datasets while they are on the platform. This ensures that sensitive data used in AI research is kept private and protected from unauthorized access. Researchers should be able to use these datasets privately, and only authorized users should have access to them.

*Admissibility (Trust)*

Furthermore, ensuring that datasets are admissible on the platform is crucial to avoid legal issues. Admissibility is a data trust property that ensures that datasets are legal and compliant with relevant laws and regulations. Researchers must ensure that they have the legal right to use the data and that it does not violate any laws or ethical guidelines. This helps to protect researchers from legal issues and ensures human safety.

B. Potential malicious users

One of the challenges in ensuring AI reproducibility is the potential for malicious users on the platform. The verification of research by other researchers is a fundamental aspect of reproducibility. However, the question arises as to who these other researchers are and whether they can be trusted. The trustworthiness of researchers on the platform is critical in preventing sensitive and powerful AI datasets from falling into the wrong hands.

*Trustworthiness (Trust)*

To prevent the misuse of AI research for cyber-attacks or other malicious activities, it is essential to establish user trust amongst researchers on the platform. Before releasing powerful AI datasets to other researchers, it is crucial to verify their identities and ensure that their actions and activities over time are consistent and reliable. Researchers must verify each other's identities before collaborating on the platform to ensure mutual trustworthiness.

Moreover, platform administrators must implement measures to prevent malicious activities, such as monitoring user activities and detecting suspicious behaviour. This will help to identify and remove any malicious users on the platform, ensuring the safety and security of the datasets and promoting ethical research practices.

*Authenticity (Trust)*

Verifying the authenticity and trustworthiness of researchers on AI reproducibility platforms is crucial in preventing the misuse of powerful datasets. Establishing user trust amongst researchers through identity verification and monitoring user activities can help to prevent malicious activities and promote ethical research practices.

C. Intellectual Theft and Competitiveness

In addition to privacy and security concerns, intellectual theft and competitiveness also pose challenges to AI reproducibility. Researchers are often reluctant to share their datasets with other researchers, fearing that they may lose control over their data, and others may gain an advantage in subsequent research. Addressing the concerns of intellectual theft and competitiveness in AI reproducibility requires providing researchers with full ownership and control over their datasets through security properties such as access control, authentication, and authorization. By providing researchers with ownership and control over their datasets, AI reproducibility platforms can foster a culture of collaboration and trust, where researchers can share their work without fear of losing control or competitive disadvantage. This will ultimately promote more ethical research practices and advance the field of AI research as a whole.

*Possession (Security)*

AI reproducibility platforms should ensure that researchers have full ownership of their datasets and have the liberty to share them with the public or keep them private. This can be achieved by implementing security properties such as access control, authentication, and authorization. Researchers should have complete control over how their datasets are used on the platform and who has access to them.

*Access Control (Security)*

Access control, for instance, allows researchers to specify who has permission to access their datasets and what actions they can perform. Access control is another essential data security property that can be implemented to ensure that only authorized users have privileges and permissions to control the use of sensitive datasets. This helps to prevent unauthorized access and misuse of the data.

*Authentication (Security)*

Authentication verifies the identity of users on the platform, while authorization ensures that only authorized users can access specific resources. This ensures that users are who they claim to be, preventing unauthorized access to data and algorithms.

*Authorization (Security)*

This ensures that users have the appropriate permissions to access and use data and algorithms, preventing unauthorized use or modification.

## 4.2 Cyber Security Properties that Ensure the AI Platforms Support Reproducibility

Victoria Stodden [12] has identified several aspects that repositories must consider to ensure reproducibility. AI platforms for AI reproducibility can utilize cyber security properties to enhance AI reproducibility while also ensuring that security measures do not hinder the reproducibility process.

For example, Stodden suggests that reproducibility can be improved by ensuring that all relevant code, data, and documentation are made available in a single location. To ensure the security of the datasets, the platform can implement confidentiality as a security property. This ensures that sensitive datasets are kept private and accessible only to authorized users, thereby protecting the data from unauthorized use or access.

Stodden also recommends that documentation is standardized, machine-readable, and in a format that can be easily processed. This can be achieved through access control, authentication, and authorization. By implementing these security properties, researchers can control the use of their datasets, ensuring that only authorized users can access and use them. This will prevent unauthorized use and promote adherence to regulations such as GDPR.

Furthermore, Stodden emphasizes the importance of version control and encourages researchers to document changes and updates made to their research. Implementing admissibility as a security property ensures that datasets are legal and meet compliance requirements. This will help researchers avoid legal issues that may arise from using illegal or non-compliant datasets.

AI platforms can utilize security properties such as confidentiality, access control, authentication, authorization, and admissibility to enhance AI reproducibility while ensuring that security measures do not hinder the process. By doing so, platforms can promote trust among researchers and ensure the integrity and confidentiality of sensitive datasets.

Confidentiality is crucial to ensuring the privacy and security of data and code analysis, which is an essential aspect of Victoria Stodden's suggested facets for reproducibility [12]. In addition to confidentiality, the ownership of artefacts is also necessary for reproducibility. Authorship plays a crucial role in reproducibility, and a security property that ensures researchers own data and code artefacts is of utmost importance. The possession and access control security properties enable researchers to own data and artefacts and control their access, promoting ownership and accountability.

In addition to the already discussed cyber security properties, the following are proposed to enhance reproducibility.

*Availability (Usability)*

In the context of AI reproducibility, availability plays a critical role in ensuring that research compendia, including data and code, are easily accessible, quickly and reliably. Availability ensures that data and algorithms used in AI research are accessible when needed, enabling the reproducibility of research results. This is important for researchers to maximize access to artefacts and for reproducibility. The need for availability is further heightened in AI research, which often involves massive amounts of data and code. To ensure availability, AI platforms must provide easy access to research compendia.

*Survivability and Responsibility (Usability)*

It is essential to consider system survivability and responsibility to ensure that datasets are preserved even if the platform crashes and the platform can handle future events. Open licensing, which provides access to researchers to reproduce other research, is also essential for availability.

*Interoperability (Usability)*

Interoperability is another critical factor for reproducible research, and usability properties can promote it. As Victoria Stodden pointed out, interoperability is necessary for reproducible research, and platforms can enhance interoperability through usability properties. Hence, to ensure AI reproducibility, platforms must prioritize availability, system survivability, responsibility, open licensing, and interoperability.

*Transparency (Trust)*

The security property of transparency is crucial to promoting reproducible research as it allows researchers to understand how datasets work and establish trust in the data and code. Metadata and documentation are also essential for this purpose.

*Validity and Integrity (Trust)*

Validity and integrity are security properties that ensure the versioning and use of persistent unique identifiers and hashes for datasets and code, which can further enhance transparency. Integrity ensures that data and algorithms used in AI research are not altered or tampered with in any way, ensuring that the results obtained are accurate and reliable. Additionally, transparency can also facilitate collaborations among researchers on the platform. Other security properties such as authenticity, accountabil-ity, reliability, and trustworthiness are also important to promote collaborations and establish trust among researchers as discussed above. Fig.1 illustrates the various security properties that can promote AI reproducibility.

*Non-repudiation (Trust)*
This ensures that users cannot deny actions taken or data used in AI research, enhancing accountability.

*Auditability (Usability)*
This ensures that all activities related to AI research are recorded and can be reviewed if necessary, providing a way to identify and address security breaches.

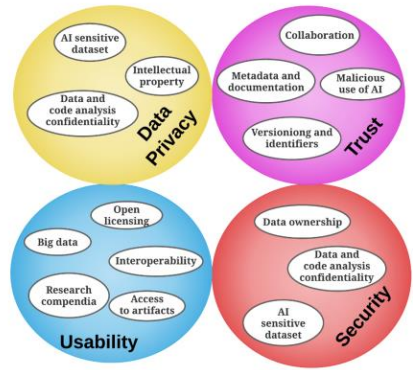

Fig. 1. Cyber security properties that enhances AI platforms for reproducibility

## 5 ANALYSIS OF THE PLATFORM FOR AI REPRODUCIBILITY

It is worth noting that the documentation provided by most AI platforms primarily focuses on using the platforms for research and reproducibility, rather than delving into security-related aspects [21]–[25]. Consequently, a substantial amount of information that could have addressed certain aspects of the framework's questions is suspected to be omitted from the documentation. For instance, the topic of authentication, a crucial aspect of security, was often not explicitly covered in the documentation. As a result, the option 'N/I' (No Information) was frequently employed to indicate the absence of specific security-related details.

The outcomes of the analysis of Floydhub, BEAT, Kaggle, Codalab, and OpenML are summarized in Table 3. In this representation, the yellow color signifies 'Yes,' the maroon color represents 'No,' and the blue color corresponds to 'N/I.' Black horizontal lines are used to separate each category, which includes data security, user security, data privacy, user privacy, data usability, user system usability, data trust, and user trust. White lines are inserted in between the three colors to distinguish one response from another. This visual representation provides a clear overview of the analysis outcomes for each cyber security property across the selected AI platforms.

TABLE 3
RESULT OF THE FIVE ANALYZED PLATFORMS (YELLOW- 'YES', MAROON- 'NO', AND BLUE- 'N/I')

| Security Properties (SPUT) | Category | Sub-factors | FloydHub | BEAT | Kaggle | Codalab | OpenML |
|---|---|---|---|---|---|---|---|
| Security | User security | Authorization | yellow | yellow | yellow | yellow | yellow |
| | | Auditability | yellow | blue | blue | blue | blue |
| | | Authentication | yellow | blue | yellow | yellow | yellow |
| | Data security | Access control | yellow | blue | yellow | yellow | maroon |
| | | Possession | yellow | yellow | yellow | yellow | yellow |
| Privacy | User privacy | Anonymity | blue | blue | blue | blue | blue |
| | | Pseudonymity | blue | blue | blue | blue | blue |
| | Data privacy | Confidentiality | yellow | maroon | yellow | yellow | yellow |
| | | Unobservability | blue | blue | blue | blue | blue |
| Usability | Data usability | Accessibility | blue | blue | blue | blue | blue |
| | | Availability | yellow | yellow | yellow | yellow | yellow |
| | | Timeliness | blue | blue | blue | blue | blue |
| | User system usability | System survivability | blue | blue | yellow | blue | blue |
| | | Intervenability | yellow | blue | yellow | blue | blue |
| | | Human safety | blue | blue | blue | blue | blue |

| | | | | | | |
|---|---|---|---|---|---|---|
| | Responsibility | | | | | |
| | Authenticity | | | | | |
| | Accountability | | | | | |
| | Consistency | | | yellow | | |
| | Non-repudiation | | | | | |
| | Trustworthiness | | | | yellow | |
| | Reliability | | | | | |
| **Data Trust** | Admissibility | | | | | yellow |
| | Accuracy | | yellow | | | yellow |
| | Integrity | | | | | |
| | Transparency | yellow | yellow | yellow | yellow | yellow |
| | Validity | yellow | yellow | yellow | yellow | yellow |

The result of the analysis highlights a notable lack of information concerning privacy in the documentation of all the analyzed AI platforms. None of the platforms appears to provide provisions for user privacy as per the information available. User privacy, in this context, pertains to concealing the identity of researchers while they are using the platforms. Furthermore, only one of the data privacy properties, namely confidentiality, is implemented by four of the platforms. Curiously, BEAT claims to have data privacy but seems to apply it exclusively to the data formats of the results, rather than the datasets themselves. In light of the proposed framework's definition, BEAT's implementation does not align with the concept of confidentiality, which revolves around researchers having access to datasets, whether they are private or public. None of the platforms seem to address unobservability, a property referring to a researcher's ability to use datasets discreetly, without others knowing they are in use, while still being accountable.

Moreover, the analysis reveals that most of the platforms lack information regarding user trust. Limited documentation exists for these properties, leading to suspicions that the platforms may not have provisions for user trust even if further investigation were conducted beyond the available documentation. Kaggle and Codalab appear to have provisions for specific aspects of user trust. Kaggle focuses on ensuring researchers' consistency in their behaviors on the platform, with a mechanism for reporting inconsistencies. Codalab introduces the concept of trustworthiness, allowing researchers to check each other's profiles to establish trust. While it is possible that other platforms might also implement trustworthiness by enabling researchers to view each other's profiles and gather basic information, this information is not present in their documentation. None of the platforms provide information for authenticity, accountability, non-repudiation, and reliability based on the analysis. These properties are intended to verify researchers' identities and hold them accountable for their actions on the platforms, ultimately fostering collaboration among researchers.

Data trust, in comparison to user trust, appears to have been implemented to some extent by most of the platforms. All the platforms implement validity through versioning of datasets, allowing users to assess the currency and reliability of datasets before using them. Transparency is implemented by four of the platforms, except for OpenML, to ensure documented information explains datasets and their usage. Admissibility is only present in OpenML, where an administrator can approve datasets before they are uploaded. Codalab is the sole platform among the five analyzed to have a provision for integrity through the use of data hashes, ensuring the originality of datasets without unauthorized modifications. Accuracy provisions are made by Codalab, BEAT, and OpenML to detect errors in datasets.

In terms of security as a property, the analyzed platforms reasonably implement possession and authorization, with all platforms assigning each uploaded dataset to an owner (possession) and enforcing rules and policies governing user actions (authorization). Authentication is implemented by four platforms, except for BEAT, which lacks information on user account creation and login. Access control is not mentioned in the documentation of BEAT and OpenML. Kaggle and Floydhub provide information about auditability, enabling the monitoring of researchers' activities on the platforms.

Lastly, none of the platforms provide information on human safety and responsibility, which are essential considerations to ensure that individuals using datasets do not encounter issues, particularly when working with sensitive data that may have legal implications due to unauthorized disclosure.

Based on the results derived from the analysis of the five AI platforms, none of these platforms appear to have implemented all the properties associated with secure systems and applications. Kaggle stands out by implementing the highest number of security properties among the five platforms, followed by Codalab, FloydHub, OpenML, and BEAT in descending order. BEAT is observed to have the lowest number of security properties implemented. Furthermore, it is apparent that all the analyzed platforms seem to provide relatively fewer provisions for trust and privacy. Table 4 offers a breakdown of the percentages of security properties provided by each platform, with the percentage calculated concerning the total number of sub-factors considered by the framework, which amounts to twenty-seven (27). These percentages provide a clear comparison of the extent to which each platform addresses security properties, highlighting the variations in their respective implementations.

TABLE 4
PERCENTAGES OF THE RESPONSES PROVIDED BY EACH PLATFORM

| AI platforms | 'Yes' | 'No' | 'N/I' |
|---|---|---|---|
| **FloydHub** | 37.0% | - | 63.0% |
| **BEAT** | 22.2% | 3.7% | 74.1% |
| **Kaggle** | 44.4% | - | 55.6% |
| **Codalab** | 40.7% | - | 59.3% |
| **OpenML** | 29.6% | 3.7% | 66.7% |

Kaggle and FloydHub have been observed to provide all the sub-factors within the categories of data and user security, which is a notable achievement. They are followed by Codalab, which has just one of the properties of user security implemented. OpenML and BEAT appear to have the lowest number of security properties in this regard.

Regarding user system usability, Kaggle emerges as the leader, offering more provisions than the other platforms. FloydHub follows with a slightly lower number, while the remaining platforms, Codalab, OpenML, and BEAT, do not seem to provide any provisions for user system usability based on the available documentation.

In terms of data trust, Codalab leads the way with the most provisions, followed by OpenML and BEAT, while Kaggle and FloydHub appear to have the lowest number of data trust properties implemented.

Table 5 provides more detailed information, offering percentages for each category provided by the platforms, shedding light on the variations in their implementations. These percentages offer a comprehensive overview of the extent to which each platform addresses specific categories of security properties, highlighting the strengths and weaknesses of their implementations in different aspects of security.

TABLE 5
PERCENTAGE OF EACH OF THE CATEGORIES PROVIDED BY THE PLATFORMS

| Cyber Security Properties (SPUT) | Category (Number of sub-factors) | FloydHub | BEAT | Kaggle | Codalab | OpenML |
|---|---|---|---|---|---|---|
| **Security** | User security (3) | 100% | 33.3% | 100% | 66.7% | 66.7% |
|  | Data security (2) | 100% | 50.0% | 100% | 100% | 50% |
| **Privacy** | User Privacy (2) | 0 | 0 | 0 | 0 | 0 |
|  | Data Privacy (2) | 50% | 0 | 50% | 50% | 50% |
| **Usability** | Data usability (3) | 33.3% | 33.3% | 33.3% | 33.3% | 33.3% |
|  | User system usability (4) | 25% | 0 | 50% | 0 | 0 |
| **Trust** | User trust (6) | 0 | 0 | 16.7% | 16.7 | 0 |
|  | Data trust (5) | 40% | 60% | 40% | 80% | 60% |

Table 5 offers insights into the strategic security focus of each platform under analysis. BEAT, despite having the fewest sub-factors overall, exhibits a notable emphasis on data trust. Kaggle, in contrast, boasts the highest number of sub-factors and places a strong emphasis on both data and user security. Kaggle also pays some attention to aspects of privacy, usability, and trust. Codalab, while addressing elements of privacy and usability to some extent, places a more significant emphasis on data and user security and data trust. OpenML's approach is akin to Codalab, with similar categories of emphasis, but it notably lacks provisions for user trust.

Kaggle, designed to accommodate both machine learning competitions and reproducibility, stands out for its comprehensive cyber security measures. Codalab, while providing similar functionalities to Kaggle, also incorporates substantial security features. Notably, both platforms benefit from a competitive context, which appears to be a driving force behind their robust security implementations. In contrast, Floydhub and BEAT, despite being recognized for their support of reproducibility by Isdahl and Gundersen [1], exhibit relatively weaker security implementations. BEAT, in particular, ranks lowest in terms of cyber security support, while Floydhub falls in the middle of the five platforms.

OpenML, not featured among the top four platforms supporting reproducibility according to Isdahl and Gundersen [1], also demonstrates a comparatively poor implementation of cyber security properties based on this research paper's analysis. Consequently, Kaggle and Codalab, previously identified as strong contenders for supporting reproducibility, emerge as the top two platforms in terms of cyber security support, underscoring their versatility in promoting both reproducibility and security.

# 6 RECOMMENDATION OF CYBER SECURITY PROPERTIES/ PLATFORMS FOR DIFFERENT USE

# CASES

The use of AI platforms for reproducibility presents a unique challenge for individual researchers, smaller laboratories, and bigger corporations. While all of these use cases require data and user security and system usability, they also have their different needs and expectations from the platforms. Individual researchers, for instance, may have limited resources and expertise, making it difficult for them to manage their datasets and perform complex analyses. As a result, they may turn to AI platforms to help them manage their data and collaborate with other researchers.

However, the use of these platforms by individual researchers raises concerns about data privacy and security. For example, some researchers may be reluctant to share their data with others due to privacy and intellectual property concerns. To address these concerns, individual researchers need AI platforms that provide robust data security and privacy features, including encryption and access control, to protect their datasets from unauthorized access.

Smaller laboratories, on the other hand, may have more resources than individual researchers but still face challenges in managing their datasets and conducting reproducible research. For instance, smaller laboratories may lack the necessary infrastructure and expertise to manage large datasets, making it difficult for them to reproduce research results. In such cases, AI platforms can provide smaller laboratories with the necessary tools and expertise to manage their data and conduct reproducible research.

To ensure data privacy and security for smaller laboratories, AI platforms must provide robust access control mechanisms and secure data storage features. These platforms must also allow smaller laboratories to maintain ownership of their data and artifacts, while at the same time promoting transparency and collaboration among researchers.

Bigger corporations may have more resources than individual researchers and smaller laboratories but also face unique challenges in managing their datasets and conducting reproducible research. For example, corporations may have multiple teams working on different projects, each with their datasets and analyses. As a result, corporations need AI platforms that provide robust access control mechanisms, data security features, and collaboration tools to ensure that all teams can access the necessary data and collaborate effectively.

To balance data privacy and security with collaboration and transparency, AI platforms for bigger corporations must provide robust access control mechanisms, secure data storage features, and transparency features such as detailed metadata and documentation. These platforms must also provide accountability and authenticity mechanisms to ensure that all actions taken on the platform are traceable and attributable to specific users

## 6.1 Individual Researcher

To provide tailored recommendations, it is important to consider the diversity of needs and expectations among individual researchers. For this purpose, we have divided them into two classes: the naïve researcher and the confident researcher.

The naïve researcher is characterized as someone who is not yet confident in their machine learning skills and is primarily seeking to learn on the AI platform away from the public eye. This group values privacy and may not be interested in visibility or collaboration at first. However, as they gain more skills and knowledge, they may become more open to collaborating and sharing their work with others.

Conversely, confident researcher is already skilled in machine learning and seeks opportunities to showcase their abilities and collaborate with others. This group values visibility and collaboration as a means to further their research and professional development.

It is important to note that both groups require data and user security and system usability on the AI platform. However, the balance of privacy and trust differs between the two groups. Naïve researchers prioritize privacy over trust as they are more concerned with learning than being visible and collaborating. On the other hand, confident researchers prioritize trust over privacy as they value visibility and collaboration to further their research and professional development.

A. Naïve researcher

The privacy and trust needs of the naïve researcher are critical and should not be overlooked. The privacy of user identity and data is of utmost importance to the naïve researcher who is more interested in learning and less interested in visibility and collaboration. Anonymity, unobservability, confidentiality, and pseudonymity are all necessary sub-factors of privacy that the naïve researcher requires. These researchers have no reason to share their identities and want to keep their activities on the platform secret. AI platforms must prioritize the implementation of privacy sub-factors to meet the needs of this group of researchers.

While trust may not be a top priority for the naïve researcher, data trust is essential. These researchers need to trust the public datasets they will be using on the platform. AI platforms must ensure the integrity of public datasets, so the naïve researcher can learn with confidence.

It is also essential that researchers are made aware of the available privacy and trust features on the platforms to enable them to make informed decisions on the use of the platforms. As asserted by [12], proper documentation and metadata are critical to promoting transparency and trust in the research community. Therefore, AI platforms need to provide adequate documentation and metadata for their privacy and trust features

None of the AI platforms adequately address the primary requirements of naïve researchers, particularly concerning user and data privacy. When it comes to data privacy, specifically the confidentiality of datasets, all the AI platforms analysed are recommended, except for BEAT, which lacks provisions for both data and user privacy.

B. Confident researcher

Confident researchers prioritize trust over privacy, as they aim to showcase their skills and collaborate with others. In particular, they require data and user trust to verify the identities of other researchers and present accurate and valid data to avoid any potential legal issues. They also need to be accountable, reliable, and consistent to attract profitable collaborations (authenticity, accountability, reliability). Additionally, transparency is essential to enhance their visibility and enable others to understand their work.

Although privacy is less of a concern for confident researchers, they may still require confidentiality to protect certain datasets from being accessed by unauthorized individuals or used for intellectual property theft.

None of the five platforms subjected to analysis comprehensively address the aspect of trust. However, with regard to data trust, Codalab receives a recommendation, boasting the highest percentage of data trust sub-factors based on our analysis. Furthermore, Codalab stands out as one of the two platforms that incorporate provisions for user trust, along with being one of the four platforms that offer confidentiality.

## 6.2 Small Laboratories

For smaller laboratories, data and user security are of great importance. They will need to ensure that their company's data and their staff's identity are kept private and secure. They will also need a platform that provides access control, allowing only authorized staff to access their company's data and artefacts.

Regarding data privacy, confidentiality is the primary sub-factor that the smaller laboratories require. They must have the liberty to choose whether to make the datasets public or private to everyone. By keeping the datasets proprietary, the smaller laboratories can protect their intellectual property and avoid potential theft or misuse of their data. They will need other privacy properties such as anonymity, and unobservability, which will enable them to work on their projects without any external intrusion.

For the trust, smaller laboratories will need properties such as authenticity and accountability, which will enable them to verify the identity of their staff and hold them responsible for their actions on the platform. They will need to ensure that their staff members are genuine and can be held accountable for their actions on the platform. Additionally, they will need a platform that provides reliable and consistent data to facilitate their research and experiments.

Transparency is not a top priority for smaller laboratories since they are not interested in visibility and external collaborations. However, metadata and documentation are essential for the lab's internal use to ensure that their staff can understand and reproduce their research and experiments.

The smaller laboratories must have a platform that provides robust security measures for user and data trust. As the laboratory team works on a company project, the platform must have user trust provisions that enable the company to monitor the activities of each staff member and ensure that they are all part of the team. The team members must be held accountable for their actions, and their identities must be authenticated and trustworthy. The datasets used for the project must also be trustworthy, accurate, and valid to avoid any issues that could arise from using erroneous or fraudulent data. Additionally, transparency is crucial to making the project visible to the team members.

None of the analyzed platforms cater adequately to the substantial trust requirements of smaller corporations. However, Kaggle and Codalab offer a glimmer of hope due to their provisions for collaboration and data confidentiality, which could potentially benefit smaller corporations seeking trust-enhancing features.

## 6.3 Big Corporations

It is important to note that big corporations have a lot to lose when it comes to data privacy breaches. Hence, their primary need is privacy, both user and data privacy. User privacy ensures that the identity of the staff working on the project is kept hidden, and their activities are not made public (anonymity, unobservability, pseudonymity). Data privacy ensures that the datasets used for the project are kept confidential and secure from external parties (confidentiality).

Although trust is also important for big corporations, privacy is their utmost priority. They have the resources to thoroughly verify the authenticity, trustworthiness, accountability, and reliability of the researchers they sponsor (user trust). They will also need to trust the datasets used for the project (data trust - admissibility, accuracy, validity, transparency).

In addition to user trust, big corporations also require data trust as they will be providing sensitive data to external researchers. They need to ensure that the data is protected and used only for the intended purposes (confidentiality). The accuracy, validity, and admissibility of the data are also crucial as the results of the research will impact their business decisions. Transparency is also necessary to monitor the progress of the research and ensure that it aligns with the company's goals and values.

However, unlike smaller laboratories, big corporations may not require as much data privacy. They may be willing to share some datasets publicly to attract top researchers and gain visibility. They may also have their own data protection measures in place, reducing their reliance on the platform's privacy features.

None of the analyzed platforms adequately address the substantial trust requirements of smaller corporations. Nevertheless, Kaggle and Codalab offer a slight possibility of meeting these trust needs due to their features related to collaboration and data confidentiality.

# 7 CONCLUSIONS AND FUTURE RESEARCH

This research has shed light on the imperative need for cyber security in platforms dedicated to enhancing AI reproducibility. It has presented a comprehensive overview of the state of security within the analyzed platforms. The demand for security in AI reproducibility has evolved into an absolute necessity. Consequently, this research offers recommendations for essential security properties that must be taken into account by the platform developers, whether in the creation of new platforms or the enhancement of existing ones. Given that many of these platforms are still in development, it is crucial to recognize the significance of security and the indispensable security properties required for AI reproducibility. Developers can leverage this research to assess the current status of their systems and applications, identifying areas where additional security measures are essential.

Furthermore, this research equips researchers with the knowledge to identify and prioritize specific security features crucial for their AI reproducibility endeavors. When faced with the task of selecting an AI platform, researchers can utilize the proposed framework for an in-depth analysis and comparison of various platforms. Its user-friendly design ensures accessibility for individuals without extensive IT training.

While the framework primarily focuses on determining whether platforms provide specific security properties, future research endeavors can delve deeper to assess the extent of implementation of each property. For instance, platforms may provide user information for verification, but the nature of this information can significantly impact the establishment of trust post-verification. Similarly, authentication measures may be in place, but the level of security provided by these measures warrants investigation. Additionally, the framework's sub-factors can be expanded, and novel security properties can be identified and incorporated. Future work may extend the analysis to encompass other AI and non-AI platforms, considering both documentation and user experience.

Notably, despite being among the AI platforms that excel in supporting reproducibility, Floydhub has ceased operations [21]. This highlights the importance of considering the financial aspects of platform sustainability in future research endeavors.

## ACKNOWLEDGMENT

I dedicate this paper to the Almighty God, my maker, Jesus Christ, my Friend and the Holy Spirit, my Help. Thank God for the inspiration and wisdom to conduct this research to His glory. I thank my ever-supporting husband (Victor Oluwasegun Falade) and children (Jedidiah and Shekinah). Thank you, Sarah Ezekiel Dangana and Terry Ezekiel Dangana, for your support.